\documentstyle[11pt,epsf,rotate]{article}

\begin{document}

\vspace*{1.5cm}

% 3/3/97
\begin{center}
{\Large\bf Interfering Doorway States and Giant Resonances.\\
 I. Resonance Spectrum and Multipole Strengths}
\\[1cm]

V.V.~Sokolov$^{1}$,
 I.~Rotter$^{2,3}$,  D.V.~Savin$^{1}$ and
M.~M\"uller$^4$
\\[8mm]
$^1$Budker Institute of Nuclear Physics, 6300090
Novosibirsk, Russia   \\
$^2$ Forschungszentrum Rossendorf, Institut f\"ur Kern- und Hadronenphysik,
\\D-01314 Dresden, Germany \\
$^3$ Technische Universit\"at Dresden, Institut f\"ur
Theoretische Physik,\\ D-01062 Dresden, Germany \\
 $^4$ Centro Internacional de Ciencias, Cuernavaca, Mexico\\

\end{center}

\vspace*{1cm}

\begin{abstract}
A phenomenological schematic model of multipole giant resonances (GR) is
considered which treats the external interaction via common decay channels
on the same footing as the coherent part of the internal residual
interaction. The damping due to the coupling to the sea of complicated
states is neglected.  As a result, the formation of GR is governed by the
interplay and competition of two kinds of collectivity, the internal and the
external one. The mixing of the doorway components of a GR due to the
external interaction influences significantly their multipole strengths,
widths and positions in energy. In particular, a narrow resonance state with
an appreciable multipole strength is formed when the doorway components
strongly overlap.
\end{abstract}

\newpage

\section[]{Introduction}
In spite of much efforts, the nature of giant resonances (GR) is only partly
understood nowadays.  It is commonly accepted that they are collective
excitations formed by coherent superpositions of many correlated
predominantly one particle -- one hole configurations with given total
quantum numbers. These superpositions are usually found by  diagonalizing
the residual interaction in the $1p-1h$ configuration space in the framework
of the random phase (RPA) or Tamm-Dancoff approximations. But their detailed
microscopic structure still needs further investigation.

The recent progress of high energy accelerators and high precision detectors
gives the possibility to analyse in coincident experiments
\cite{OES82,EFH84,FEH85,BBD87,AYD94} the direct particle decays of the giant
resonance states into specific channels. The decay rates extracted from the
data contain the desired microscopic information. They are therefore a very
useful tool for a careful study of the collective modes of nuclear motion.

At the GR energies, the value which is typical for the escape widths of the
configurations contributing to the collective mode exceeds usually their
level spacings. The energy continuum effects play therefore an important
role and cannot be treated as a perturbation slightly broadening  the levels
which, nevertheless, remain isolated.  A few methods exist to take into
account the energy continuum in a straightforward manner.  Some of them
\cite{SB75,GEPFS81,CMOS82,HKS82,CK85,GNU86} extend the RPA by including
exactly the single particle continuum.  The more general approach
\cite{BKM69,baroho78,hobaro79,KW84} based on the explicit separation of the
intrinsic and channel subspaces is closely related to Feshbach's idea of
doorway states \cite{FKL67} which we exploit in the present paper.

Generally, a giant resonance consists of one or a few doorway states on the
background of many complicated fine structure states. Even when within the
RPA the corresponding doorway states are well isolated from one another,
their overlapping with the background states can give rise to interference
effects \cite{AHKS72,KR86} which cannot be described by a sum of independent
Breit-Wigner contributions. However, due to the nearly chaotic structure of
the background states, the main  effect is \cite{SZ90,ZB91,MU94}, after
energy averaging, the damping of the doorway states described by the
spreading width.  In contrast, the interference of the overlapping doorway
states with each other may significantly influence the form of the energy
spectrum of the decay products of giant resonances as shown in
\cite{SZ90,ZB91}.

In this paper we investigate the interference effects caused by the
overlapping of doorway  components of giant resonances.  We use the
extension, proposed in \cite{SZ90}, of the simple Brown-Bolsterli
\cite{BB59} schematic model for multipole collective nuclear excitations to
open (decaying) systems. Although being qualitative, this phenomenological
model still maintains the main features of the real situation. The giant
resonances emerge out of the interplay between two different kinds of
collective behaviour: the synchronized collective intrinsic motion and the
cooperative particle emission.

The present part I of our paper is devoted to the consideration of the
collective excitation modes of the internal motion in an unstable nucleus
while in the following part II \cite{two} the cross section pattern is
discussed.

Our schematic model along with its formal solution are presented in sect. 2.
The two kinds of collectivity are analysed analytically in sect. 3 while
their interplay and the interference of overlapping doorway resonances are
described in sect. 4. Here, an important connection between the resonance
spectrum and the multipole strengths of the doorway states is established.

The transition strengths as well as the escape widths  get significantly
redistributed between the overlapping doorway states when the interaction
via the energy continuum becomes strong. The GR is mainly formed by two
doorway states which acquire comparable parts of the total multipole
strength. One of these states turns out to be much narrower than the other
one. Both are lying at energies which are lower than the
energy  at which the giant resonance would lie if the mixing via the
continuum would be absent.

In sect. 5, we show numerical results obtained in the same model with and
without the restrictions introduced into the preceding analytical study. The
results  confirm the main features of the interference picture.  At strong
external coupling, they can be qualitatively understood in the two-level
approximation introduced in sect. 4. The results are summarized in sect. 6.

We use the matrix shorthands throughout the paper. The capital letters are
used for matrices in the Hilbert space of the internal motion; matrices in
the space of the scattering channels are marked by the hat symbol. The {\it
column} vectors in the internal space as well as the {\it row} vectors in
the channel one are represented by bold letters.

\setcounter{equation}{0}
\section[]{The Model}
As usual (see for example \cite{MW69}), we suggest that a hierarchy of
complexities of the internal states of the system under consideration
exists. The first class contains the simplest states presumably of
$1p-1h$ nature which are directly connected to the continuum by
appreciable transition matrix elements $A_n^c$. The states of the other
classes of complexity have no direct connection to the continuum, at least
to the same set of channel states as the first class.  They can decay
into these channels only through the states of the first class
 to which they are connected by some residual internal interaction.

In the  present part of the paper, we  restrict ourselves to the
consideration of the dynamics of the states of the first class.  The role of
the background of complicated states will be discussed in the second part  of
our paper \cite{two}, refered in what follows as part II.  As in the
pioneering work by Brown and Bolsterli \cite{BB59}, we choose the Hamiltonian
of the internal motion  in the subspace of the states of the class 1 in the
form
\begin{equation}
\label{Hp} H = H_0 + {\bf D}{\bf D}^T\;.
 \end{equation}
It consists of the unperturbed part $H_0$ containing N discrete intrinsic
levels $e_n (n=1,2,...,N$), and of a factorized residual interaction of,
let us say, dipole-dipole type.  Besides by the internal residual
interaction, the intrinsic levels are mixed also by the external interaction
via common decay channels \cite{F58,MW69,KobNO69,KirNO69}. This interaction,
\begin{equation}\label{W}
W = AA^T \qquad \Longrightarrow \qquad W_{mn} = \sum_{c(open)} A_m^c A_n^c\;,
\end{equation}
originates from on-shell self-energy contributions of all open channels.
The matrix $A$ consists of $k$
$N$-dimensional column vectors ${\bf A}^c$ connecting all internal states
with each channel $c$. These vectors are real because of time-reversal
invariance. In the following we neglect a possible smooth energy dependence
of the components $A_n^c$ over the whole energy domain considered.  The
validity of such an assumption is not always obvious and deserves a special
consideration.

Adding to $H$ the external interaction (\ref{W}) via common decay channels,
we get the nonhermitian effective Hamiltonian
\begin{equation}\label{meffH}
{\cal H} = H_0 + {\cal H}^{(int)} = H_0 + {\bf D}{\bf D}^T -
\frac{i}{2}\,AA^T\equiv H -\frac{i}{2}\,W
\equiv {\cal H}_0 + {\bf D}{\bf D}^T
\end{equation}
which describes giant resonances as appearing out of the interplay of two
kinds of collectivity, the internal and the external one. The expression
(\ref{meffH}) is the many-channel version of the effective Hamiltonian
considered in reference  \cite{SZ90}. Having omitted the coupling to the sea
of complicated states, we neglect in particular the internal damping and
spreading widths of the collective excitations formed by superpositions of
the states of the first class which are embedded into the continuum. The
coupling to the states of other classes will be considered in part II  of
our paper \cite{two}  by using statistical methods.

One finds due to the factorized form of the internal residual interaction
\begin{equation}\label{Det}
{\rm Det}\left({\cal E} - {\cal
H}\right) = {\rm Det}\left({\cal E} - {\cal H}_0\right)\,
\left[1 - {\bf D}^T\frac{1}{{\cal E}-{\cal H}_0}{\bf D}\right]\;.
\end{equation}
Therefore, the spectrum of unstable (resonance) states is given in our
model by the equation
\begin{equation}\label{Seq}
1 - {\bf D}^T\frac{1}{{\cal E}-{\cal H}_0}{\bf D} =
1 - {\bf D}^T\,{\cal G}_0({\cal E}){\bf D} = 0
\end{equation}
in the complex energy plane.

The factorized structure of the external interaction (\ref{W}) allows us
to simplify the Green's matrix ${\cal G}_0({\cal E})=
({\cal E}-{\cal H}_0)^{-1}$ to
\begin{equation}\label{CalG_0}
{\cal G}_0({\cal E}) = G_0({\cal E}) - \frac{i}{2}\,G_0({\cal E})\,A\,
\frac{1}{1+\frac{i}{2}{\hat K}_0({\cal E})}\,A^T\,G_0({\cal E})
\end{equation}
by subsequent iterations of the Dyson equation
\begin{equation}\label{Dyeq_0}
{\cal G}_0({\cal E}) = G_0({\cal E}) -
\frac{i}{2}\,G_0({\cal E})\,W\,{\cal G}_0({\cal E})\;.
\end{equation}
Here $G_0({\cal E})=({\cal E}-H_0)^{-1}$ is the unperturbed Green's matrix
while ${\hat K}_0({\cal E})=A^TG_0({\cal E})A$ is the corresponding
$K$-matrix. These matrices describe the unperturbed intrinsic motion and its
coupling to the continuum, respectively. Therefore they have the levels
$e_n$ as their poles in the complex energy plane.

With the help of (\ref{CalG_0}), equation (\ref{Seq}) can be presented
in the form
\begin{equation}\label{rSeq}
1 - {\bf D}^T\,G_0({\cal E})\,{\bf D} + \frac{i}{2}\,{\bf D}^T\,G_0({\cal E})
\,A\,\frac{1}{1+\frac{i}{2}{\hat K}_0({\cal E})}\,A^T\,
G_0({\cal E})\,{\bf D} = 0\;.
\end{equation}
The last term on the l.h.s. describes the influence of the external
interaction via the continuum onto the energy spectrum of the internal
motion. This equation can easily be reduced to an algebraic equation of the
order $N$ producing the $N$ complex energies of the resonance states.

In a similar manner we obtain
\begin{equation}\label{mRm}
{\hat K}(E) = A^T\,\frac{1}{E-H}\,A =
{\hat K}_0(E) + A^T\,G_0(E)\,{\bf D}\,
\frac{1}{1-{\bf D}^T\,G_0(E)\,{\bf D}}\,{\bf D}^T\,G_0(E)\,A
\end{equation}
for the full $K$-matrix. The additional term is due to the internal
interaction.  Eqs.  (\ref{rSeq}) and (\ref{mRm}) present the  explicit
solution of our model. In the following we investigate this solution
analytically as well as numerically.
Before finishing this section, let us consider the limit of a very strong
internal interaction (see below for the explicit condition). It is well
known that such an interaction leads to the creation of a collective
vibration mode which is shifted in energy from the location of the original
unperturbed (parental) levels by a distance of the order of magnitude of
$Tr({\bf D}{\bf D}^T)={\bf D}^2$. Characterizing this location by some
average position $\varepsilon_0 = \langle e \rangle$, we find
\begin{equation}\label{Cegr}
{\cal E}_{gr} = E_{gr} - \frac{i}{2}\,\Gamma_{gr} =
\varepsilon_0 + {\bf D}^2 -
\frac{i}{2}\,{\hat {\bf A}}_d^2
\end{equation}
from the secular equation (\ref{rSeq}) for the complex energy of the only
isolated giant resonance state. In the energy domain of this state, the
$K$-matrix (\ref{mRm}) acquires the standard resonance form
\begin{equation}\label{K_gr}
{\hat K}(E) = \frac{{\hat {\bf A}_d^T}{\hat {\bf A}_d}}{E-E_{gr}}
\end{equation}
where the components of the row vector ${\hat {\bf A}}$ are equal to the
projections
\begin{equation}\label{A_d}
A_d^c\equiv ({\bf d}\cdot{\bf A}^c)
\end{equation}
of the decay amplitudes ${\bf A}^c$ onto the unit dipole vector ${\bf d}=
{\bf D}/\sqrt{{\bf D}^2}$. The residues
\begin{equation}\label{mPw}
\Gamma_{gr}^c = \left({\bf d}\cdot{\bf A}^c\right)^2
\end{equation}
of the diagonal elements of the $K$-matrix (\ref{K_gr}) determine the partial
escape widths of the GR state.

Because of the overlapping of different resonance states, eqs. (\ref{rSeq}\,,
\ref{mRm}) lead, generally,  to a more complicated picture. The elements of
the channel space matrix ${\hat K}(E)$ can be presented by
\begin{equation}\label{TrfR}
K^{c\,c'}(E) = {\rm Tr}\,\left(G(E)\,{\bf A}^c\,({\bf A}^{c'})^T \right)
\end{equation}
as the matrix trace in the Hilbert space of the internal motion. Therefore,
it is invariant with respect to any transformation of the basis in this
space. Such transformations connect different parametrizations of the
$K$-matrix. The concrete choice of the parametrization is dictated by
physical reasons as well as by convenience. For instance, one can use the
eigenbasis of the hamiltonian $H$ to represent the $K$-matrix as the sum
\begin{equation}\label{Rmd}
{\hat K}(E) = \sum_r\frac{{\hat {\bf A}}_r^T{\hat {\bf A}}_r}
{E-\varepsilon_r}
\end{equation}
over all the internal eigenstates ${\bf\Phi}^{(r)}$. Each term of this
sum is directly analogous to the single-resonance expression (\ref{K_gr}).
The row vectors ${\hat{\bf A}_r}$ consist of the real components
\begin{equation}\label{A_r}
A_r^c = {\bf\Phi}^{(r)}\cdot{\bf A}^c
\end{equation}
along the eigenvectors ${\bf\Phi}^{(r)}$ belonging to the eigenenergies
$\varepsilon_r$. The residues
\begin{equation}\label{Rpw}
\Gamma_r^c = (A_r^c)^2
\end{equation}
at the poles of the diagonal elements of the matrix (\ref{Rmd}) characterize
the coupling of the intrinsic eigenstates ${\bf \Phi}^{(r)}$ to individual
channel states. In analogy with (\ref{mPw}),
 we will call them the
partial escape widths of the $r$th resonance. It must be stressed however
that, contrary to the case of isolated resonances, the real residues
$\Gamma_r^c$ differ from the  generally complex residues at the poles of
the scattering amplitudes when the resonances overlap.  For this reason,
the very concept of the partial widths becomes ambiguous when overlapping
resonances are concerned (see part II  \cite{two}  for more details).

\setcounter{equation}{0}
\section[]{Collective Phenomena}
\subsection{Internal Collectivity}
It is convenient to treat the factorized internal residual interaction
${\bf D}{\bf D}^T$ formally in the same manner as the external one introducing
an additional imaginary "channel" with the "decay amplitudes"
${\bf A}^0\equiv \sqrt{2i}{\bf D}$. Defining the new matrix
\begin{equation}\label{enA}
{\cal A} = \left({\bf A}^0\;\;{\bf A}^1\;\;.\;.\;.\;\;\;{\bf A}^k\right)\;,
\end{equation}
we will consider the matrix
\begin{equation}\label{enR}
{\hat {\cal K}}(E) = {\cal A}^T\,G(E)\,{\cal A}
\end{equation}
in the enlarged channel space. Besides the $k\times k$ block ${\hat K}(E)$
this matrix contains along the main diagonal additionally the function
\begin{equation}\label{enR00}
{\cal K}^{0\,0}(E)\equiv 2i\,P(E) = 2i\;{\bf D}^T\,G(E)\,{\bf D}
\end{equation}
which carries information on the degree of collectivity of the internal
motion.

The degree of collectivity of an internal eigenstate ${\bf \Phi}^{(r)}$ of
the hermitian part $H$ of the effective Hamiltonian (\ref{meffH}) is
characterized by the residue $\left({\bf \Phi}^{(r)}\cdot {\bf D}\right)^2$
of the function $P(E)$ at its pole at the eigenenergy $E=\varepsilon_r$. To
estimate this degree, let us note first that the collectivity can become
appreciable only if the internal interaction is sufficiently strong and
dominates the internal dynamics. Under such a condition it is natural to
start with the diagonalization of the interaction matrix ${\bf D}{\bf D}^T$.
Because of its factorized structure, this matrix possesses the only nonzero
eigenvalue ${\bf D}^2$ belonging to the eigenstate
\begin{equation}\label{Ups1}
{\bf \Upsilon}^{(1)} = {\bf d}\;.
\end{equation}
The rest of the eigenstates ${\bf \Upsilon}^{(\mu)},\;\mu =2,3,...,N$
remains arbitrary because of the degeneracy of the zero eigenvalue.  These
states span a basis in the $(N-1)$-dimensional subspace orthogonal to the
vector ${\bf d}$. We will fix this basis a little bit later.

In the new basis the function $P(E)$ reduces to
\begin{equation}\label{PUps}
P(E) = {\rm Tr}\left(\frac{1}{E-H}\,{\bf D}{\bf D}^T\right) =
{\bf D}^2\,G_{coll}(E)
\end{equation}
where $G_{coll}(E)$ is the upper diagonal matrix element of the internal
Green's matrix. The Hamiltonian matrix $H$ looks as follows:
\begin{equation}\label{HUps}
H = \left( \begin{array}{cc}
\varepsilon_0 + {\bf D}^2 & {\bf h}^T \\
{\bf h} & {\tilde H}
\end{array}
\right)\;.
\end{equation}
Here the energy
\begin{equation}\label{mep}
\varepsilon_0 \equiv \sum_n e_n\,d_n^2  = \langle e \rangle
\end{equation}
is the weighted mean position of the parental levels $e_n$. The
$(N-1)$-dimensional vector ${\bf h}$ has the components
\begin{equation}\label{dUps}
h_{\mu} = \sum_n e_n\,d_n\,\Upsilon_n^{(\mu)}
\end{equation}
while the matrix elements of the $(N-1)\times (N-1)$ submatrix $\tilde H$ are
equal to
\begin{equation}\label{tilH}
{\tilde H}_{\mu\,\nu} =
\sum_n e_n\,\Upsilon_n^{(\mu)}\,\Upsilon_n^{(\nu)}\;.
\end{equation}

Representing similar to (\ref{HUps}) the resolvent $G(E)$ as
\begin{equation}\label{blG}
G(E) = \left( \begin{array}{cc}
G_{coll}(E) & {\bf F}^T(E) \\
{\bf F}(E) & {\tilde G}(E)
\end{array}
\right)\;,
\end{equation}
one finds
\begin{equation}\label{Gcoll}
G_{coll}(E) = \frac{1}{E - \varepsilon_0 - {\bf D}^2 -
{\bf h}^T\,\frac{1}{E-{\tilde H}}\,{\bf h}}
\end{equation}
when
\begin{equation}\label{F}
{\bf F}(E) = \frac{1}{E-{\tilde H}}\,{\bf h}\;\,G_{coll}(E)
\end{equation}
and
\begin{equation}\label{tilG}
{\tilde G}(E) = \frac{1}{E-{\tilde H}} +
\frac{1}{E-{\tilde H}}\;{\bf h}\,{\bf h}^T\,\frac{1}{E-{\tilde H}}\;\,
G_{coll}(E)\;.
\end{equation}

Each eigenvalue $\varepsilon_r$ of the Hamiltonian $H$ satisfies the equation
\begin{equation}\label{lmbd}
\lambda(E)\equiv E - \varepsilon_0 - {\bf D}^2 -
{\bf h}^T\,\frac{1}{E-{\tilde H}}\,{\bf h} = 0
\end{equation}
and the value
\begin{equation}\label{str}
f^r = \left({\bf d}\cdot{\bf\Phi}^{(r)}\right)^2 =
Res P(\varepsilon_r)/{\bf D}^2 =
\left(\frac{d\lambda(E)}{dE}\right)_{E=\varepsilon_r}^{-1} =
\left[1+{\bf h}^T\,\left(\frac{1}{\varepsilon_r -
{\tilde H}}\right)^2\,{\bf h}\right]^{-1}\;,
\end{equation}
subject to the condition
\begin{equation}\label{sumf}
\sum_r f^r = 1\;,
\end{equation}
describes the part of the dipole strength carried by the eigenstate
${\bf \Phi}^{(r)}$.

Further, we diagonalize the submatrix (\ref{tilH}),
\begin{equation}\label{dtilH}
\sum_n e_n\,\Upsilon_n^{(\mu)}\,\Upsilon_n^{(\nu)} =
\tilde\varepsilon_{\mu}\,\delta_{\mu\,\nu}
\end{equation}
by choosing the till now  unspecified basic vectors ${\bf \Upsilon^{(\mu)}}$
to coincide with its eigenvectors. Using the completeness condition
\begin{equation}\label{Ccon}
\sum_{\nu}\Upsilon_m^{(\nu)}\Upsilon_n^{(\nu)} = \delta_{m\,n} -
 d_m\,d_n
\end{equation}
one easily obtains
\begin{equation}\label{eigUps}
{\bf \Upsilon}^{(\mu)} =
-h_{\mu}\,\frac{1}{{\tilde\varepsilon}_{\mu}-H_0}\,{\bf d}\;.
\end{equation}
The matrix elements $h_{\mu}$ play the role of the normalization coefficients
of the eigenvectors and are equal to
\begin{equation}\label{hmu}
h_{\mu} = \left[{\bf d}^T\left(\frac{1}{{\tilde\varepsilon}_{\mu}-
H_0}\right)^2{\bf d}\right]^{-\frac{1}{2}}\;.
\end{equation}

The orthogonality condition ${\bf d}\cdot{\bf \Upsilon}^{(\mu)}=0$
immediately leads to the equation
\begin{equation}\label{tilep}
{\bf d}^T\,\frac{1}{{\tilde\varepsilon}-H_0}\,{\bf d} =
\sum_n\frac{d_n^2}{{\tilde\varepsilon}-e_n} = 0
\end{equation}
for the eigenvalue spectrum of the submatrix $\tilde H$. Obviously, each
eigenvalue ${\tilde\varepsilon_{\mu}}$ lies between two neighboring parental
levels $e$. Therefore, the levels $\tilde\varepsilon$ are shifted, with
respect to the original ones, by distances of the order of magnitude of the
unperturbed mean level spacing. This is much smaller than the energy shift
$\sim {\bf D}^2$ of the collective level in the upper corner of the
Hamiltonian matrix (\ref{HUps}).

The collective level $\varepsilon_{coll}=\varepsilon_0+{\bf D}^2$ is still
mixed with the $N-1$ levels ${\tilde\varepsilon_{\mu}}$ via the matrix
elements $h_{\mu}$. The square length of the vector ${\bf h}$ turns out to
be equal to the variance
\begin{equation}\label{sqh}
{\bf h}^2 = \sum_n e_n^2\,d_n^2 - \left(\sum_n e_n d_n^2\right)^2 =
\sum_n(e_n-\langle e\rangle)^2\,d_n^2 = \langle(e-\langle e\rangle)^2\rangle
=\Delta_e^2
\end{equation}
of the distribution of the parental levels. This leads to the estimation
$|h_{\mu}|\sim \Delta_e/\sqrt{N-1}\;$ of the individual matrix elements.
Therefore, the mixing is governed by the parameter $\kappa=\Delta_e/
{\bf D}^2\,$. Suggesting that this parameter is small, $\kappa\ll 1$, one
can use the standard perturbation expansion which gives
\begin{equation}\label{epsr}
\varepsilon_1 = \varepsilon_0 + \left(1+\kappa^2\right)\,{\bf D}^2\;,
\qquad
|\varepsilon_r - \tilde{\varepsilon}_r|\sim \frac{\kappa^2}{N-1}\,{\bf D}^2
\;\;\;(r\neq 1)
\end{equation}
for the levels and
\begin{equation}\label{apstr}
f^1 = 1-\kappa^2\;, \qquad f^r \sim \frac{\kappa^2}{N-1}\;\;\; (r\neq 1)
\end{equation}
for the dipole strengths. The first level accumulates the lion's share of
both the total dipole strength and the energy displacement. In the limit
$\kappa \rightarrow 0$ the collectivity of the first level becomes perfect
while the rest of the levels carries no collectivity at all.

\subsection[]{External Collectivity}
Let us now turn to the properties of the $K$-matrix eq. (\ref{mRm}).
The strong internal interaction causes a remarkable redistribution of
the original  residues $\Gamma^c_n=\left(A_n^c\right)^2$.  In this
case, the $\Upsilon$-basis (\ref{Ups1},\,\ref{eigUps}) constructed above
becomes a preferential one. Taking into account eqs. (\ref{F},\,\ref{tilG})
one gets in this basis
\begin{equation}\label{UpsR}
{\hat K}(E) =
\left[{\hat {\bf A}_d} +
{\bf h}^T\,\frac{1}{E-{\tilde H}}\,A_{\bot}\right]^T\,
\left[{\hat {\bf A}_d} +
{\bf h}^T\,\frac{1}{E-{\tilde H}}\,A_{\bot}\right]\;G_{coll}(E) +
A_{\bot}^T\,\frac{1}{E-{\tilde H}}\,A_{\bot}\;.
\end{equation}
The rectangular submatrix $A_{\bot}$ is composed of $(N-1)$-dimensional
column vectors ${\bf A}_{\bot}^c$ orthogonal to the dipole vector,
\begin{equation}\label{Aorth}
\left({\bf d}\cdot{\bf A}^c_{\bot}\right) = 0\,, \qquad
A_{\mu}^c = \left({\bf \Upsilon}^{(\mu)}\cdot{\bf A}^c\right),
\end{equation}
whereas the row vector ${\hat {\bf A}_d}$ of the longitudinal components
$A_d^c$ is defined in (\ref{A_d}). It can easily be checked that the
contributions of the poles at the energies $E={\tilde\varepsilon_{\mu}}$
in the two terms on the r.h.s. perfectly cancel each other. The actual poles
of the $K$-matrix are given by the roots $\varepsilon_r$ of the equation
(\ref{lmbd}).

It immediately follows from (\ref{UpsR}) that  the partial widths
(\ref{Rpw}) are equal to
\begin{equation}\label{UpsRpw}
\Gamma_r^c = f^r\left[A_d^c +
{\bf h}^T\,\frac{1}{\varepsilon_r-{\tilde H}}
\,{\bf A}_{\bot}^c
\right]^2
\end{equation}
and depend on the relative strength $\kappa$ of the residual mixing. In
particular, by using condition (\ref{Ccon}), one finds
\begin{equation}\label{est1}
\biggl|{\bf h}^T\,\frac{1}{\varepsilon_1-{\tilde H}}\,{\bf A}_{\bot}^c\biggr|
\approx\frac{1}{{\bf D}^2}\,|({\bf h}\cdot{\bf A}_{\bot}^c)|\approx
\frac{1}{{\bf D}^2}\,\Bigl|\sum_n(e_n-
\varepsilon_0)d_n\left(A_{\bot}^c\right)_n\Bigr|\leq
\kappa\,\sqrt{\left({\bf A_{\bot}}^c\right)^2}
\end{equation}
for the collective level if $\kappa\ll 1$.

In the square bracket of eq. (\ref{UpsRpw}) the first term dominates for the
collective level $r=1$ as long as $|A_d^c|/|A^c_{\bot}|\gg \kappa$. Therefore,
$\Gamma_1^c\approx \left(A_d^c\right)^2$ under such a condition. The
remaining part $\left({\bf A}_{\bot}^c\right)^2$ is distributed over the
$N-1$ levels lying in the energy interval $\sim\Delta_e$ around the point
$\varepsilon_0$. In this region the pattern turns out to depend crucially on
the ratio $\left({\bf A}_{\bot}^c\right)^2/\Delta_e^2$. Each state acquires
the partial width $\sim\left({\bf A}_{\bot}^c\right)^2/(N-1)$ if this ratio
is small while a strong redistribution of the widths occurs in the opposite
case $\left({\bf A}_{\bot}^c\right)^2/\Delta_e^2\gg 1$. It is called "width
collectivization" \cite{SZ88,SZ89,SZ92} or "trapping effect"
\cite{ro91,isrodi,hebrro,ismuro,sodiro,chem}: $k$ eigenstates ${\bf
\Phi}^{(r)}$ get large components along the vector ${\bf A}_{\bot}^c$ and
accumulate almost the total value $\left({\bf A}_{\bot}^c\right)^2$ (see
also \cite{PV88,RMPD90,someda}). This phenomenon was first observed in
realistic numerical simulations of nuclear reactions in
\cite{Ml68,KR85,R88}.

In the limit $\kappa=0$, when the internal collectivity is maximal, the
expression (\ref{UpsR}) reduces to
\begin{equation}\label{UpsRtl}
{\hat K}(E) =
\frac{{\hat {\bf A}_d}^T\,{\hat {\bf A}_d}}{E-\varepsilon_{coll}} +
\frac{{\hat X}_{\bot}}{E-\varepsilon_0}\;.
\end{equation}
Here the matrix
\begin{equation}\label{Xorth}
{\hat X}_{\bot} = \hat X - {\hat {\bf A}_d}^T\,{\hat {\bf A}_d} =
A_{\bot}^T\,A_{\bot}
\end{equation}
is composed of the scalar products $({\bf A}_{\bot}^c\cdot
{\bf A}_{\bot}^{c'})$ in the orthogonal subspace when the matrix
\begin{equation}\label{X}
{\hat X} = A^T\,A
\end{equation}
is formed by the scalar products $\left({\bf A}^{(c)}\cdot
{\bf A}^{(c')}\right)$ \cite{M68,SZ88} in the full Hilbert space.

One can immediately see from eq. (\ref{UpsRtl}) that the partial widths
(\ref{Rpw}) of the only collective state with the energy
$\varepsilon_{coll}=\varepsilon_0+{\bf D}^2$
 are equal to:
\begin{equation}\label{Rpwcoll}
\Gamma_{coll}^c = \left(A_d^c\right)^2
\end{equation}
in agreement with eq. ({\ref{mPw}}).

In the given limit, the other states are jointly presented in (\ref{UpsRtl})
by the single pole at the energy $E=\varepsilon_0$. The residues at this
pole do not factorize contrary to the residue at the pole of the collective
part. This means that different linear superpositions of the original states
are excited via different channels at the same energy $\varepsilon_0$.  To
find the
corresponding partial widths $\Gamma_r^c$,
(\ref{Rpw}), one has first to
diagonalize the matrix ${\hat X}_{\bot}$. Then
\begin{equation}\label{Rpwsin}
\Gamma_r^c =
\gamma_{\bot}^r\left({{\xi_{\bot}}_c^{(r)}}\right)^2\,,
\qquad (r=1,2,...,k)
\end{equation}
 is expressed in terms of the eigenvalues $\gamma_{\bot}^r$ and the
left eigenvectors $\xi_{\bot}^{(r)}$ of the matrix ${\hat X}_{\bot}$. One
sees that in the considered limit of very strong internal collectivity only
$k$ superpositions out of the $N-1$ ones with the energy $\varepsilon_0$
possess nonzero partial width (\ref{Rpw}).
They absorb the part
\begin{equation}\label{Rpwsinsum}
\sum_{r=1}^{r=k}\Gamma_r^c = \left({\bf A}_{\bot}^c\right)^2 = ({\bf
A}^c)^2 - \Gamma^c_{coll}
\end{equation}
of the total original value $({\bf A}^c)^2$. We conclude that the partial
widths $\Gamma_r^c$ of the $k+1$ states presented in the $K$-matrix
(\ref{UpsRtl}) are formed by contributions of all parental states.

\setcounter{equation}{0}
\section[]{Interplay of Two Kinds of Collectivity. Interference of Doorway
Resonances}
\subsection{The Doorway Basis}
Now we turn to  the resonance spectrum resulting from the interplay and
competition of both kinds of collectivity. Since the interaction plays the
dominant role in the dynamics studied, we start with the consideration of
the interaction matrix
\begin{equation}\label{efint}
{\cal H}^{(int)} = {\bf D}{\bf D}^T - \frac{i}{2}\,AA^T\;.
\end{equation}
The manifold of the $k+1$ linearly independent vectors ${\bf D}$ and
${\bf A}^c$ forms a $(k+1)$-dimensional subspace in the internal Hilbert
space the total dimension of which is $N$. It is convenient to choose the
first $k+1$ basis vectors of the total Hilbert space in such a manner that
they belong entirely to this subspace. Then only the upper left $(k+1)
\times (k+1)$ block of the interaction matrix will contain non-zero matrix
elements. We proceed in the following three steps:

(i) Let us first orthogonalize the set of the $k$ vectors ${\bf A}^c$. For
this purpose,  we diagonalize the matrix $\hat X$, eq. (\ref{X}), of the
scalar products of these vectors. Let ${\hat\xi}$ be the matrix of the (left)
eigenvectors,
\begin{equation}\label{xi}
{\hat\xi}\,{\hat X}={\hat \gamma}\,{\hat\xi}
\end{equation}
where
\begin{equation}\label{gm}
\hat\gamma = diag\,(\gamma^1\;\gamma^2\;.\;.\;.\;\gamma^k)
\end{equation}
is the diagonal matrix of the eigenvalues. It is then obvious that the
rectangular matrix
\begin{equation}\label{ma}
a = A{\hat\xi}^T\gamma^{-\frac{1}{2}}
\end{equation}
consists of $k$ mutually orthogonal unit vectors
\begin{equation}\label{va}
{\bf a}^c=\frac{1}{\sqrt{\gamma^c}}\,\sum_{c'}\xi_{c'}^{(c)}\,{\bf A}^{c'}\;.
\end{equation}
Adding to this set an extra unit vector ${\bf a}^0$ which is orthogonal to
all of them, one obtains a new basis in the non-trivial part of the total
Hilbert space. One can easily see that the vectors ${\bf a}^0,\,{\bf a}^c$
are just the eigenvectors of the antihermitian part $W=A\,A^T$. Therefore,
this matrix becomes diagonal,
\begin{equation}\label{dW}
W = diag\,(0\;\;\hat\gamma)\;.
\end{equation}
Its nonzero eigenvalues $\gamma^c$ coincide with those of the matrix $\hat
X$ \cite{SZ88,SZ89}. For the present, we drop the matrix blocks and the
vector components which belong to the complementary $(N-(k+1))$-dimensional
subspace and consist of zero elements.

In the chosen basis the unit dipole vector ${\bf d}$ has the components
$$d_0 = ({\bf a}^0\cdot{\bf d}) = {\rm sin}\Theta\,,\qquad
d_c = ({\bf a}^c\cdot{\bf d}) = {\rm cos}\Theta\,{\rm cos}\varphi_c\,;$$
\begin{equation}\label{Da}
\sum_c{\rm cos}^2\varphi_c=1\;.
\end{equation}
Here we have introduced the angle $\Theta, (0\leq\Theta\leq\pi/2)$, between
the dipole vector ${\bf D}$ and the $k$-dimensional subspace spanned by the
decay vectors ${\bf A}^c$. This angle is an important parameter which
governs the interference effects under consideration.

The matrix of the internal interaction reads
\begin{equation}\label{DDa}
{\bf D}^2\,\left( \begin{array}{cc}
{\rm sin}^2\Theta &
{\rm sin}\Theta\,{\rm cos}\Theta\;{\bf l}^T\\
{\rm sin}\Theta\,{\rm cos}\Theta\;{\bf l} &
{\rm cos}^2\Theta\;{\bf l}\,{\bf l}^T
\end{array} \right)
\end{equation}
where ${\bf l}$ stands for the unit vector with the components
$l_c={\rm cos}\varphi_c$.

(ii) Next we  rotate the considered $(k+1)$-dimensional subspace around
the unit vector ${\bf a}^0$ to put the unit dipole vector ${\bf d}$ into
the coordinate plane $(0,1)$. This is done by the diagonalization of the
$k\times k$ submatrix ${\bf l}\,{\bf l}^T$ with the help of a $k$-dimensional
orthogonal matrix $\;{\hat\eta}\;$ the first column of which coincides with
the vector ${\bf l}$. This transformation resembles that described in
subsection 4.1.

Now only two nonzero components of the unit dipole vector ${\bf d}$
are left which are equal to
\begin{equation}\label{dvec}
d_0 = {\rm sin}\Theta\,,\qquad d_1 = {\rm cos}\Theta\;,
\end{equation}
and only the $2\times 2$ upper block of the internal interaction matrix
(\ref{DDa}) remains non-trivial. The vectors ${\bf A}^c$ are transformed into
\begin{equation}\label{Avec}
A_0^c = 0\,,\qquad A_1^c =
\sum_{c'}\sqrt{\gamma^{c'}}\,{\rm cos}\varphi_{c'}\,\xi_c^{(c')}\,,
\qquad A_{\alpha}^c =
\sum_{c'}\sqrt{\gamma^{c'}}\,\eta_{c'}^{(\alpha)}\,\xi_c^{(c')}\;,
\end{equation}
so that
\begin{equation}\label{A_1d}
A_d^c = {\rm cos}\Theta\,A_1^c =
{\rm cos}\Theta\,
\sum_{c'}\sqrt{\gamma^{c'}}\,{\rm cos}\varphi_{c'}\,\xi_c^{(c')}
\end{equation}
when the lower diagonal submatrix $\hat\gamma$ in eq. (\ref{dW}) is replaced
by
\begin{equation}\label{Wtr}
{\hat\gamma}\rightarrow \left( \begin{array}{cc}
\langle\gamma\rangle & {\bf w}^T \\
{\bf w} & \tilde W
\end{array} \right)\;.
\end{equation}
Here
\begin{equation}\label{strW}
\langle\gamma\rangle = \sum_c\left(A^c_1\right)^2 =
\sum_c\gamma^c\,{\rm cos}^2\varphi_c\,,\qquad
w^{(\alpha)} = \sum_c\gamma^c\,{\rm cos}\varphi_c\,\eta_c^{(\alpha)}\,,
\;\;\;(\alpha=2,3,...,k)
\end{equation}
and
\begin{equation}\label{subW}
W_{\alpha\,\alpha'} = \sum_c\gamma^c\,\eta_c^{(\alpha)}\,\eta_c^{(\alpha')}\;.
\end{equation}

(iii). Returning now into the total $N$-dimensional Hilbert space, the two
consecutive transformations just described are part of the global
transformation produced by the orthogonal matrix
\begin{equation}\label{Om}
\Omega =\left({\bf\Omega}^{(0)}={\bf a}^{(0)}\;\;\;\,
{\bf\Omega}^{(1)}=\sum_c{\rm cos}\varphi_c\,{\bf a}^{(c)}\;\;\;\,
{\bf\Omega}^{(2\leq\alpha\leq k)}=\sum_c\eta_c^{(\alpha)}\,{\bf
a}^{(c)}\;\; \;\,{\bf \Omega}^{(k+1\leq s\leq N-1)}\right)\;.
\end{equation}
In (\ref{Om}), the two groups of vectors, the $N - (k+1)$ vectors
${\bf \Omega}^{(k+1\leq s\leq N-1)}$ in the full space and the $k-1$ ones
$\eta^{(2\leq s=\alpha\leq k)}$ in the $k+1$ - dimensional subspace, can
still be chosen arbitrarily. We will fix them later.

In the ${\bf\Omega}$-basis, the diagonal matrix elements of the unperturbed
Hamiltonian $H_0$ are given by the weighted mean positions
\begin{equation}\label{tile}
{\tilde e}_s = \sum_n e_n \left(\Omega_n^{(s)}\right)^2, \qquad
(s=0,1,...,N-1)
\end{equation}
when the off-diagonal elements
\begin{equation}\label{Vss'}
V_{ss'}  = \sum_n e_n\,\Omega_n^{(s)}\,\Omega_n^{(s')}\,,\qquad (s\neq s')
\end{equation}
obey the general sum rules
\begin{equation}\label{conV}
\sum_{s'\neq s}V_{ss'}^2=\sum_{s'}(e_{s'}-{\tilde e}_s)^2\,
\left(\Omega_n^{(s)}\right)^2\sim\Delta_e^2
\end{equation}
(compare with eq. (\ref{sqh})). Since we do not expect any special relation
between the original basis and the doorway one, all off-diagonal matrix
elements are suggested to be of the same order of magnitude. This leads to
the estimation
\begin{equation}\label{estVss'}
\big|V_{s\neq s'}\big|\sim \Delta_e/\sqrt{N-1}
\end{equation}
similar to that found in subsection 4.1.

We now use the last $N-(k+1)$ vectors ${\bf \Omega}^{(s)}\;,
(s\equiv tr=k+1,k+2,...,N-1)$ in (\ref{Om}) in order to diagonalize the
lower block of the unperturbed Hamiltonian \cite{SZ89,SZ92},
\begin{equation}\label{tr}
\sum_n e_n\,\Omega_n^{(tr)}\,\Omega_n^{(tr')} =
\tilde\varepsilon_{tr}\,\delta_{tr\,tr'}\;.
\end{equation}
In the new picture, the $N_{tr}=N-k-1$ eigenstates with the energies
$\tilde\varepsilon_{tr}$ (which lie within the original energy region
$\Delta_e$ \cite{SZ92}) are "trapped" \cite{R88,SZ88,PV88}, i.e. they do not
have a direct access to the continuum. These states can decay only via the
first $N_{dw}=k+1$ "doorway" states  to which they are coupled
by the hermitian residual interaction
\begin{equation}\label{Vtrdw}
V_{dw\;tr} = \sum_n e_n\,\Omega_n^{(dw)}\,\Omega_n^{(tr)}\,.
\end{equation}
This interaction appears from the initial unperturbed hamiltonian being
transformed into the doorway basis.

A typical value for the widths of the doorway  states is
$\langle\gamma\rangle\approx\frac{1}{k}\,Tr\,W\sim\langle\left({\bf A}^c
\right)^2\rangle$. Only one of them can, under certain conditions, become
almost stable (see subsection 5.2) but then it is displaced by a distance
$\sim{\bf D}^2$. Therefore,  all trapped states lie in the complex
energy plane far from the $N_{dw}$ doorway states and their admixture
to the latter is small as one of the ratios
\begin{equation}\label{kk'}
\kappa = \frac{\Delta_e}{{\bf D}^2}\;,\qquad
\kappa' = \frac{\Delta_e}{\langle\gamma\rangle}\;.
\end{equation}
According to the estimation (\ref{estVss'}) the trapped states acquire the
widths $\sim\frac{N_{dw}}{N_{tr}}\,{\kappa'}^2\langle\gamma\rangle$. The
energy shifts of the trapped states are of the same order of magnitude. These
states are responsible therefore for the fine structure effects (with the
characteristic energy scale $\Delta_e/(N_{tr})$) in the energy domain of the
parental levels. They become irrelevant when $(\kappa\,,\kappa')
\rightarrow 0$.

The doorway $N_{dw}\times{N_{dw}}$ part of the effective Hamiltonian
\begin{equation}\label{effHdw}
{\cal H}^{(dw)} = \left( \begin{array}{cc}
{\cal H}^{(coll)} & \chi^T \\
\chi & \tilde{\cal H}
\end{array} \right)
\end{equation}
includes two different blocks along the main diagonal. Only the upper
$2\times 2$ block
\begin{equation}\label{effHcoll}
{\cal H}^{(coll)} = \left( \begin{array}{cc}
{\tilde e}_0 + {\rm sin}^2\Theta\,{\bf D}^2 &
V_{01} + {\rm sin}\Theta{\rm cos}\Theta\,{\bf D}^2 \\
V_{10} + {\rm sin}\Theta{\rm cos}\Theta\,{\bf D}^2 &
{\tilde e}_1 + {\rm cos}^2\Theta\,{\bf D}^2
\end{array} \right)
-\frac{i}{2}\langle\gamma\rangle\left( \begin{array}{cc}
0 & 0 \\ 0 & 1
\end{array} \right)
\end{equation}
contains, along with the collectivity via the continuum, the collective
effects induced by the internal residual interaction. Due to the mixing
described by the off-diagonal matrix elements, the widths, energy shifts
and dipole strengths of the two eigenstates of this block are generally
comparable to each other provided that the angle $\Theta$ differs from $0$
and $\pi/2$ so that ${\rm sin}^2\Theta\sim{\rm cos}^2\Theta$.

The $(k-1)\times (k-1)$ block
\begin{equation}\label{subeffH}
\tilde{\cal H}_{\alpha\,\alpha'} = \sum_{c,c'}\eta_c^{(\alpha)}\,
\left[
\sum_n e_n\,a_n^{(c)}\,a_n^{(c')} -\frac{i}{2}\,\gamma^c\,\delta_{c\,c'}
\right]\,\eta_{c'}^{(\alpha')}
\end{equation}
(see eq. (\ref{subW}) for the antihermitian part) describes the states
without both dipole strengths (according to (\ref{dvec})) and collective
energy shifts. Generally, they are strongly mixed with each other. The set
of vectors $\eta^{(\alpha)}$ can be used to diagonalize either the hermitian
or the antihermitian part of ${\tilde{\cal H}}$ depending on which of them
dominates the dynamics inside this block. The remaining part may then be
treated as a weak perturbation. However when  both, the hermitian and
antihermitian parts, are of equal importance the full Hamiltonian
${\tilde{\cal H}}$ must  be diagonalized.

The two doorway blocks just described are coupled by the complex interaction
\begin{equation}\label{Chi}
\chi = \left({\bf v}^{(0)}\;{\bf v}^{(1)}\right) -
\frac{i}{2}\,\left({\bf 0}\;{\bf w}\right)\;;
\end{equation}
\begin{equation}\label{meChi}
v_{\alpha}^{(0)}\equiv V_{0\alpha}\;,\qquad
v_{\alpha}^{(1)}\equiv V_{1\alpha}\;,\qquad
w^{(\alpha)} = \sum_c\gamma^c\,{\rm cos}\varphi_c\,\eta_c^{(\alpha)}\;.
\end{equation}
The influence of the hermitian part is weak as $\kappa^2,\kappa'^2$ again.
The strength of the antihermitian coupling can be estimated by using the
identity
\begin{equation}\label{sqw}
{\bf w}^2\equiv \sum_{\alpha}\left(w^{(\alpha)}\right)^2 =
\sum_c\left(\gamma^c\right)^2 {\rm cos}^2\varphi_c -
\left(\sum_c\gamma^c\,{\rm cos}^2\varphi_c\right)^2 =
\Delta_{\gamma}^2
\end{equation}
which is the counterpart of the eq. (\ref{sqh}).  In the limit $\kappa^2,
\kappa'^2\rightarrow 0$, the two blocks of doorway states with and without
internal collectivity are mixed only due to the antihermitian interaction
via the continuum. The strength of this interaction is determined by the
variance of the nonzero eigenvalues of the antihermitian part $W$, eq.
(\ref{W}).

\subsection[]{Resonance Spectrum and Dipole Strengths\\ of Doorway  States}

The interaction $W$ of the intrinsic states via the continuum causes a
strong redistribution of the dipole strength when the doorway states
overlap.  In this case, the dipole strengths of the decaying states rather
than those of the intrinsic eigenvectors ${\bf \Phi}^{(r)}$ (as defined in
(\ref{str})) should be used to characterize electromagnetic properties of
the open system considered. The natural and appropriate extension of the
definition (\ref{str}) to the dipole strength of an eigenvector ${\bf
\Psi}^{(s)}$ of the total nonhermitian effective Hamiltonian (\ref{meffH})
is given by
\begin{equation}\label{resstr}
{\tilde f}^s = \frac{1}{U_s}\Big|{\bf d}\cdot{\bf\Psi}^{(s)}\Big|^2
\end{equation}
where $U_s=\left({{\bf\Psi}^{(s)}}^*\cdot{\bf\bf\Psi}^{(s)}\right)$.
It is in accordance with the definition of expectation values if the
Hamiltonian is non-hermitean and the eigenfunctions form a bi-orthogonal
system \cite{baroho78}.

The quantities (\ref{resstr}) are directly linked with the resonance spectrum.
Multiplying the equation
\begin{equation}\label{Psi}
{\cal H}\Psi = \Psi{\cal E}
\end{equation}
for the matrix $\Psi$ of the unstable eigenstates ${\bf \Psi}^{(s)}$ by
the matrix $\Psi^{\dag}$ from the left side and adding the hermitian
conjugate of the relation thus received, one obtains
\begin{equation}\label{f-sp}
{\tilde f}^s = \frac{1}{{\bf D}^2}\left(E_s-\sum_n e_n
\frac{|\Psi_n^{(s)}|^2}{U_s}\right)
\end{equation}
where $E_s$ is the resonance energy.

In the doorway basis introduced in subsection 5.1, the upper $N_{dw}\times
N_{dw}$  block of the effective Hamiltonian is totally decoupled from the
lower block of the trapped states if one neglects the small matrix elements
(\ref{Vtrdw}) the contributions of which are of the order of magnitude
$\kappa^2$ or $\kappa'^2$. Omitting them, one neglects the fine structure
variations of the transition amplitudes in the energy domain of the parental
levels as mentioned above (sec. 5.1;  see also part II). In such an
approximation, the trapped states remain stable and are entirely excluded
from all further calculations. Then relation (\ref{f-sp}) becomes especially
simple
\begin{equation}\label{appf-sp}
{\tilde f}^{dw} = \frac{1}{{\bf D}^2}\left(E_{dw} - \varepsilon_0\right)\;.
\end{equation}

Taking into account eq. (\ref{effHdw}), the secular equation (\ref{Seq})
can now be reduced to
\begin{equation}\label{Seqdw}
{\rm Det}\left({\cal E} - {\cal H}^{(coll)} - {\cal Q}({\cal E})\right) = 0\;.
\end{equation}
The second order self-energy matrix
\begin{equation}\label{Q}
{\cal Q}({\cal E}) = \chi^T\,\frac{1}{{\cal E}-\tilde{\cal H}}\,\chi
\end{equation}
describes the  virtual transitions between the two types of doorway states.
Its explicit form depends on the interference regime inside the second group.
We further assume that the antihermitian part dominates in the Hamiltonian
submatrix (\ref{subeffH}). Therefore we diagonalize first this part by
demanding the vectors $\eta^{(\alpha)}$ to satisfy the conditions
\begin{equation}\label{dtilW}
W_{\alpha\,\alpha'} = \sum_c\gamma^c\,\eta_c^{(\alpha)}\,\eta_c^{(\alpha')}
= \tilde{\gamma}^{\alpha}\delta_{\alpha\alpha'}\;.
\end{equation}
The opposite case with dominating hermitian part can be treated in an
analogous manner.

Going further along the same line as in subsection 4.1, one finds for the
eigenvectors
\begin{equation}\label{eigeta}
\eta_c^{(\alpha)} = w^{(\alpha)}\frac{{\rm cos}\varphi_c}
{\gamma^c-\tilde{\gamma}^{\alpha}}
\end{equation}
with the normalization condition
\begin{equation}\label{norw}
w^{(\alpha)} = \left[\sum_c\frac{{\rm cos}^2\varphi_c}
{\left(\gamma^c-\tilde{\gamma}^{\alpha}\right)^2}\right]^{-\frac{1}{2}}\;.
\end{equation}
The corresponding eigenvalues $\tilde{\gamma}^{\alpha}$ are the roots of the
equation
\begin{equation}\label{tilg}
\sum_c\frac{{\rm cos}^2\varphi_c}{\gamma^c-\tilde\gamma} = 0\;.
\end{equation}
Each of the $k-1$ eigenvalues $\tilde{\gamma}^{\alpha}$ lies between two
neighbouring values $\gamma^c$. The last three equations should be compared
with eqs. (\ref{eigUps}) - (\ref{tilep}).

Considering the hermitian part of the hamiltonian (\ref{subeffH}) to be a
weak perturbation, one obtains
\begin{equation}\label{tilEn}
\tilde{\cal H}_{\alpha\alpha'}\approx \left(\tilde{e}_{\alpha} -
\frac{i}{2}\,\tilde{\gamma}^{\alpha}\right)\,\delta_{\alpha\alpha'} =
\tilde{\cal E}_{\alpha}\,\delta_{\alpha\alpha'}
\end{equation}
in first approximation. The corrections are proportional to the ratio
$\Delta_e^2/\Delta_{\gamma}^2$ of the variances of the unperturbed levels
$e_n$ and of the collective widths $\gamma^c$.
The approximation is justified when this ratio is small.

Under the last condition, one can also neglect the hermitian part of the
coupling matrix $\chi$, eq. (\ref{Chi}). The only nonzero matrix element
in the right lower corner of the self-energy matrix ${\cal Q}$ reads then
\begin{equation}\label{Q11}
{\cal Q}_{11}(E) = - \frac{1}{4}\,\sum_{\alpha}
\frac{{w^{(\alpha)}}^2}{E-\tilde{\cal E}_{\alpha}}\equiv
-\frac{1}{4}\,q(E)
\end{equation}
and the secular equation (\ref{Seqdw}) reduces to
\begin{equation}\label{appSeqdw}
\Lambda({\cal E})\equiv \left({\cal E}-\varepsilon_0\right)\,
\left({\cal E}-\varepsilon_{coll}\right)
+\frac{i}{2}\omega({\cal E})\,\left({\cal E}-\varepsilon_0-
{\rm sin}^2\Theta\,{\bf D}^2\right) = 0
\end{equation}
where the notation
\begin{equation}\label{om(E)}
\omega({\cal E}) = \langle\gamma\rangle - \frac{i}{2}\,q({\cal E})
\end{equation}
has been introduced. Here we neglected the matrix elements $V_{01}=V_{10}$
and set ${\tilde e}_0={\tilde e}_1=\varepsilon_0$. The corresponding
corrections are again proportional to $\kappa^2\,,{\kappa'}^2$. The equation
(\ref{appSeqdw}) is equivalent to an algebraic equation of $(k+1)$th order.
It determines the complex energies of the $k+1$ doorway resonances.

\subsection[]{The Two-Level Approximation}
Let us temporarily omit also the second term $q(E)$ in eq. (\ref{om(E)}).
Then the secular equation (\ref{appSeqdw}) reduces to the same quadratic one
which appears in the single-channel problem investigated in \cite{SZ90}.
In this approximation, one is left with two interfering collective levels
only.  The latter problem can be easily solved exactly. (See for example
\cite{SB94} and \cite{mudiisro} where different aspects of the problem are
treated. It has much in common with the physics of the text-book systems of
the neutral kaons \cite{KobNO69,O82}, the $\rho$ and $\omega$ mesons
\cite{F72} or the $2^+$ doublet in $^{8}Be$ \cite{KirNO69,Br92}.) Using the
notation
\begin{equation}\label{z}
z = \frac{{\cal E}-\varepsilon_0}{{\bf D}^2}\,,
\end{equation}
one obtains explicitly
\begin{equation}\label{z_0,1}
z_0 = \frac{1}{2}(1-|x|) - \frac{i}{2}\;\lambda\,\frac{1}{2}(1-|y|)\,,\qquad
z_1 = \frac{1}{2}(1+|x|) - \frac{i}{2}\;\lambda\,\frac{1}{2}(1+|y|)
\end{equation}
where
\begin{equation}\label{x}
|x| = \frac{1}{\sqrt{2}}\left[\sqrt{\left(1-\frac{1}{4}\lambda^2\right)^2 +
\lambda^2{\rm cos}^2 2\Theta} + \left(1-
\frac{1}{4}\lambda^2\right)\right]^{\frac{1}{2}}\leq 1
\end{equation}
and
\begin{equation}\label{y}
|y| = \frac{1}{\sqrt{2}}\left[\sqrt{\left(1-
\frac{4}{\lambda^2}\right)^2 +
4\;\frac{4}{\lambda^2}\,{\rm cos}^2 2\Theta} + \left(1-
\frac{4}{\lambda^2}\right)\right]^{\frac{1}{2}}\leq 1\;.
\end{equation}
Apart from the angle $\Theta$, the interference of the collective states
depends on the ratio
\begin{equation}\label{lam}
\lambda\equiv \frac{\langle\gamma\rangle}{{\bf D}^2}
\end{equation}
of the strengths of the external and internal interactions. The solution
(\ref{z_0,1}) is valid when $0 <\Theta < \pi/4$; for $\pi/4 <\Theta < \pi/2$
the imaginary parts of the two roots are to be replaced by each other. To be
definite, we consider the first possibility below.

The quantity $|x|$ measures the energy distance between the two resonances,
\begin{equation}\label{disE}
E_1 - E_0 = |x|\,{\bf D}^2   \; ,
\end{equation}
whereas $|y|$ measures the difference of their total widths,
\begin{equation}\label{difG}
\Big|\Gamma_1-\Gamma_0\Big| = |y|\,\langle\gamma\rangle\;.
\end{equation}
According to (\ref{appf-sp},\,\ref{x}), one further has
\begin{equation}\label{app0f-sp}
{\tilde f}^{0,1} = \frac{1}{2}(1\mp |x|) = {\rm Re}\,z_{0,1}
\end{equation}
in the same approximation. The latter expression shows that the closer the
resonances are to each other the more similar are their dipole strengths.

The situation is especially simple for the angle $\Theta=\pi/4$. In this case
\begin{equation}\label{pi/4}
|x| = \left\{ \begin{array}{cc}
\sqrt{1-\lambda^2/4}, & \lambda<2 \\
0\,, & \lambda>2
\end{array} \right.
; \qquad
|y| = \left\{ \begin{array}[c]{cc}
0\,, & \lambda<2 \\
\sqrt{1-4/\lambda^2} & \lambda>2
\end{array} \right.\;.
\end{equation}
In the limit $\lambda\ll 2$, the two collective levels $dw=0$ and $dw=1$ are
separated by a large distance $\sim {\bf D}^2$ but have the same widths.
The level $dw=1$ carries the whole dipole strength. With growing $\lambda$
the levels are getting closer and finally merge when $\lambda$
reaches the value 2. For $\lambda >2$, the dipole strengths as well as the
energies of both resonances remain equal to each other while their widths
differ more and more with increasing $\lambda$.

The transition at the point $\lambda=2$ gets smoother for other values of
the angle $\Theta$ but still exists as long as $\Theta$ is not too
close to 0 or $\pi/2$. If the internal interaction prevails and
$\lambda\ll 2$, both collective levels have comparable widths,
\begin{equation}\label{lm<1}
{\cal E}_0 = \varepsilon_0 - \frac{i}{2}\,{\rm sin}^2\Theta\,
\langle\gamma\rangle\;,\qquad
{\cal E}_1\equiv {\cal E}_{gr} = \varepsilon_0 + {\bf D}^2 -
\frac{i}{2}\,{\rm cos}^2\Theta\,\langle\gamma\rangle\;,
\end{equation}
while only the second one is displaced
by the distance ${\bf D}^2$ and carries the whole dipole strength. According
to eqs. (\ref{dvec}, \ref{Avec}), the total width  of this level is equal to
\begin{equation}\label{grwidth}
\Gamma_1 = {\rm cos}^2\Theta\,\langle\gamma\rangle = {\hat {\bf A}_d^2} =
\Gamma_{gr}
\end{equation}
in full agreement with eq. (\ref{Cegr}). In the opposite case of the
dominating external coupling, $\lambda\gg 2$, the energy displacement
${\bf D}^2$ is shared by the two collective resonances,
\begin{equation}\label{lm>1}
{\cal E}_0 = \varepsilon_0 + {\rm sin}^2\Theta\,{\bf D}^2
-\frac{i}{2}\langle\gamma\rangle\,\frac{1}{\lambda^2}{\rm sin}^2 2\Theta
\;,\qquad
{\cal E}_1 = \varepsilon_0 + {\rm cos}^2\Theta\,{\bf D}^2
-\frac{i}{2}\langle\gamma\rangle\left(1-\frac{1}{\lambda^2}
{\rm sin}^2 2\Theta\right)\;.
\end{equation}
(Here we omitted the small corrections $\sim\lambda^{-2}$ to the positions
of the resonances). The corresponding dipole strengths are, in this case,
equal to
\begin{equation}\label{0,1str}
{\tilde f}^0\approx {\rm sin}^2\Theta\,,\qquad
{\tilde f}^1\approx {\rm cos}^2\Theta\;.
\end{equation}
The nucleon width of the level $dw=0$ decreases with growing $\lambda$.
Finally this resonance practically disappeares from the decay spectrum in
the particle channels. In agreement with eq. (\ref{0,1str}) it gets however
a nonvanishing radiation width and contributes in the photoemission process
(see  part II).

Let us now consider the role of the other doorway resonances
$\tilde{\cal E}_{\alpha}\approx\varepsilon_0 - \frac{i}{2}
\langle\gamma\rangle$, eq. (\ref{tilEn}). Substituting the complex energies
${\cal E}_{0,1}$ found above into the sum (\ref{Q11}), one sees that the
denominators of the terms of this sum contain, as a rule, one of the large
quantities ${\bf D}^2$ or $\langle\gamma\rangle$. According to eq.
(\ref{sqw}), such terms are of the order of magnitude $\tau^2/k$ or
$\tau'^2/k$ where the parameters
\begin{equation}\label{tau}
\tau = \frac{\Delta_{\gamma}}{{\bf D}^2}\,,\qquad
\tau' = \frac{\Delta_{\gamma}}{\langle\gamma\rangle}
\end{equation}
may be expected to be reasonably small. Therefore, these doorway states
acquire a relatively small dipole strength. The interference of the first
two collective states remains most important and the above two-resonance
approximation gives a description which is at least qualitatively
satisfactory. When, however, some of the levels ${\tilde{\cal E}}_{\alpha}$
fall by chance anomolously close to one of the former two, these doorway
states take part in the interference and the picture becomes more complicated.

\setcounter{equation}{0}
\section[]{Numerical Results and Discussion}

To simplify the analytical study, we restricted ourselves in the foregoing
sections to the case of a very strong interaction ${\cal H}^{(int)} $
(\ref{efint}) so that the energy range $\Delta_e$ of the unperturbed levels
$e_n$ could be neglected. In such an approximation, the $N_{tr}=N-k-1$
trapped states are almost fully decoupled from the continuum  and only
the $N_{dw}=k+1$ collective doorway states remain relevant. In this
limit, the  resulting picture is determined essentially by the ratio
$\lambda=\langle\gamma\rangle /{\bf D}^2$ of the strengths of external and
internal interactions and by the angle $\Theta$ between the dipole vector
${\bf D}$ and the $k$-dimensional subspace of decay vectors ${\bf A}^c$.

Assuming further that all the vectors  ${\bf A}^c$ are pairwise orthogonal
and have the same lengths,  the matrix ${\hat X}$ of the scalar products
becomes proportional to unity and the problem is reduced  exactly to
the case  considered in subsection 4.3 with only two decaying states
which are mixed and share the total dipole strength. The  physics
arising from the interference of two resonance states is governed by the
effect of avoided resonance crossing. According to eq.  (\ref{appf-sp}), the
dipole strengths of the two resonances behave very much like their
positions in energy  when considered as functions of $\lambda$.  While the
sums of the two resonance energies and of the two strengths remain constant,
the corresponding differences decrease as functions of increasing $\lambda$
up to certain minimum values which depend on the angle $\Theta$ only. The
widths of the two states increase first with increasing $\lambda$ but
bifurcate for large $\lambda$.

The results of this two level approximation are illustrated in Fig.1.
The angle $\Theta$ is chosen to give cos$^2\Theta\approx 0.65$ and
$\varepsilon_0$ is set to zero.  The energies of the two collective
resonances (measured in units of the total energy displacement ${\bf D}^2$)
and  their dipole strengths plotted in dependence on $\lambda$ in Fig. 1(a)
coincide perfectly. Fig.1(b) displays the behaviour of the widths of the two
resonance states versus their energies and/or dipole strengths when
$\lambda$ changes in the interval $0\div 5$.

In the following, we  lift the above mentioned simplifications and
check the relevance of the
obtained analytical results by performing numerical
calculations under less restrictive assumptions.  We have chosen $N=10$
levels $e_n$ distributed more or less homogeneously and coupled to $k=3$
open particle decay channels. The extension of the parental spectrum of the
$N$ discrete levels $e_n$ is from $-0.2$ to $0.2$ in relative units of the
total energy displacement ${\bf D}^2$. This implies that $\kappa\equiv
\Delta_e/{\bf D}^2 \approx 0.4$.  As in Fig. 1, we set $\Theta\approx
36.3^{\circ}$ but the lengths of the vectors ${\bf A}^c$ differ from one
another within 10\%. The angles $\theta_{cc'}$ between the pairs  ${\bf
A}^c$ and ${\bf A}^{c'}$ are confined to $0.17\le |\cos\theta_{cc'}|\le
0.31$.

In  Fig.2, the energies and dipole strengths of all $10$  resonances are
plotted as a function of $\lambda$ while the changes of their total
widths with $\lambda$ are shown in the representation of the $\Gamma_s$
versus the positions $E_s$ and dipole strengths ${\tilde f}^s$, respectively.
For small $\lambda$, there is only one displaced state the dipole strength of
which is very close to unity.  With $\lambda$ increasing, first $N_{dw}=4$
doorway states  appear three of which are formed according to the three open
decay channels from the group of $N-1$ states lying around $E=0$, while the
fourth state with large dipole strength lies at the energy $E/{\bf
D}^2\approx 1$. These four states almost exhaust the whole sum of widths
$Tr\,W$ and the total dipole strength while the internal as well as external
collectivity of the $N_{tr}=6$ trapped states remain small.  Fig. 2(b) shows
that the total dipole strength is distributed mainly over two states: the
original dipole state and one out of the group around $E=0$. It must be
noted that while in the limit $\kappa=0$ the sum rule $\sum_s {\tilde f}^s
=1$ for the dipole strengths defined as in eqs. (\ref{resstr},\ref{f-sp}) is
fulfiled rigorously, this is not exactly the case in general. Indeed, the
sum $\sum_s{\tilde \varepsilon_s}$ of all mean positions
$$\tilde{\varepsilon}_s = \sum_n\frac{|\Psi_n^{(s)}|^2}{U_s}\,e_n$$
of the parental levels can slightly differ from $Tr\,H_0=\sum_n\,e_n$ which
leads to a violation of the sum rule. The strenghts of the trapped states
are not perfectly controlled by this rule.  However, Fig. 3 convinces one
that this violation remains weak in the whole region of $\lambda$
considered.

The width of the  doorway state which acquires with growing $\lambda$ the
main part of the dipole strength lost by the original collective excitation
is smaller than those of the other three broad states.  Finally, it will be
trapped at very large $\lambda$ (Fig. 2(c)). This behavior of the two states
is qualitatively quite similar to that in the two-level approximation
(compare Fig. 1). However, an appreciable part of the dipole strength is
 moved to the other low-lying doorway components which will not
be trapped.

Thus, the numerical results confirm the qualitative picture of the interplay
of the two kinds of collectivity by which the giant resonance excitations
are formed. The coherent internal dipole-dipole residual interaction
together with the external interaction  via common decay channels creates a
concentration of both, the dipole strength and the escape width, on a few
collective doorway states.

\section[]{Summary}

Summarizing, we state the following.  On the basis of a phenomenological
schematic model we have investigated  the formation of overlapping
doorway components of a giant multipole resonance. The interplay and
competition of the two kinds of collective behavior induced by the internal
and external coupling, respectively, give rise to a nontrivial interference
between these components. Two very different energy scales are formed due to
the internal dipole-dipole interaction: In the limit of zero coupling to the
continuum all levels with the exception of the collective one are confined
to the energy interval $\Delta_e$ while the  latter is displaced far away by
the distance ${\bf D}^2\gg \Delta_e$. With increasing external interaction
via $k$ open common decay channels, the width collectivization takes place
if $TrW$ exceeds the interval $\Delta_e$. This happens when the main
overlapping parameter $\lambda=\langle\gamma\rangle/ {\bf D}^2$ is still
small. As a result, $k+1$ states get escape widths being comparable to one
another while $N-k-1$ states become trapped. The $k+1$ states absorbing the
total width $TrW$ are the collective doorway states.

When $TrW$ approaches the value ${\bf D}^2$, with further increasing
$\lambda$ a second stage begins: the widths are redistributed once more
being accompanied this time by a strong redistribution of the dipole
strength and an energy shift of mainly two doorway states. The width of one
of these doorway states starts to decrease and it becomes finally trapped in
the limit of very large $\lambda$. As a result, a narrow state with a
large multipole moment is created due to the external interaction.

The internal damping of the collective motion due to the coupling to
complicated compound states has been omitted at this stage. It will be taken
into account in part II of this paper where we will also study the influence
of the interferences discussed in this paper onto the cross section in order
to allow a qualitative comparison with experimental data.  It will be
shown there that the radiation of the nearly trapped doorway state gives,
under certain conditions, the only visible contribution to the $\gamma$-ray
emission from the giant resonance.

\vspace{1cm}

{\small
\noindent
{\bf Acknowledgment:}
We are grateful to E. Kolomeizew and  E. Persson for their interest in this
work.  One of us (V.V.S.) thanks V.G. Zelevinsky for discussion of the
results. The present investigations are supported by the Deut\-sche
Forschungsgemeinschaft (Ro 922/1,6), by the  Grant  94-2058 from INTAS and
by the Deutscher Akademischer Austauschdienst.

%\newpage

\newpage

%\subsection*{Figure captions}

%\noindent {\bf Fig.1}   \\
%The $\lambda$-dependence of resonance energies and dipole
%strengths (a) and the logarithm of the widths versus energies or strengths (b)
%with $\lambda$ varying  from 0 to 5  in the two-level approximation.
%\\

%\noindent {\bf Fig.2}     \\
%The $\lambda$-dependence of resonance energies (a) and dipole strengths (b).
%The logarithm of the widths versus energies (c) and dipole strengths (d) with
%$\lambda$ varying from 0 to 5 in steps of 0.02. (Inset of (d) shows the
%magnified region of small ${\tilde f}^{s}$ in the double log-scale). For
%parameters see text.
%\\

%\noindent {\bf Fig.3}   \\
%The sum of dipole strengths eqs. (\ref{resstr}, \ref{f-sp}) as a function
%of $\lambda$.  
%\\

\begin{figure}[t]
\unitlength 1cm
\begin{picture}(20,5)
\epsfxsize 21cm
\put(-2,-9){\rotate[r]{\epsfbox{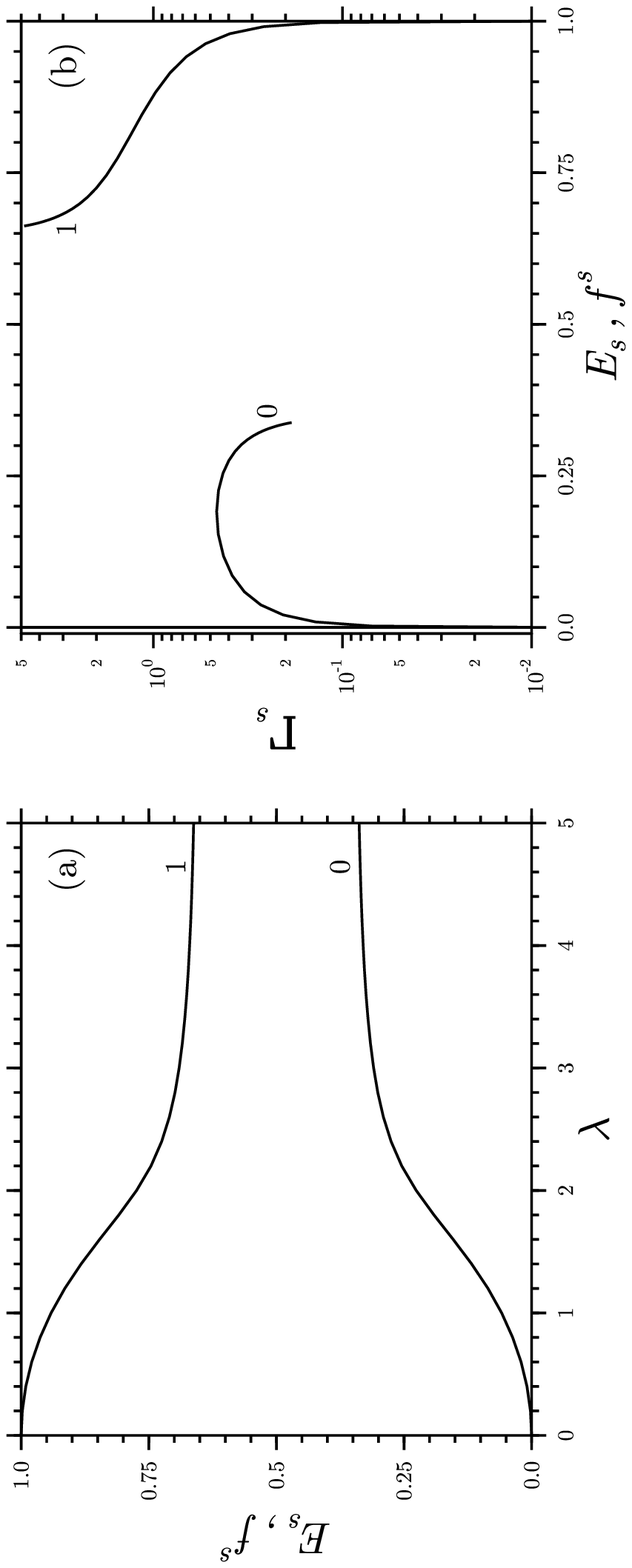}}}
\end{picture}
\unitlength 1bp
\caption{
The $\lambda$-dependence of resonance energies and dipole
strengths (a) and the logarithm of the widths versus energies or strengths (b)
with $\lambda$ varying  from 0 to 5  in the two-level approximation.
}
%\label{fig1}
\end{figure}

\begin{figure}[t]
\unitlength 1cm
\begin{picture}(20,15)
\epsfxsize 21cm
\put(-2,-6){\rotate[r]{\epsfbox{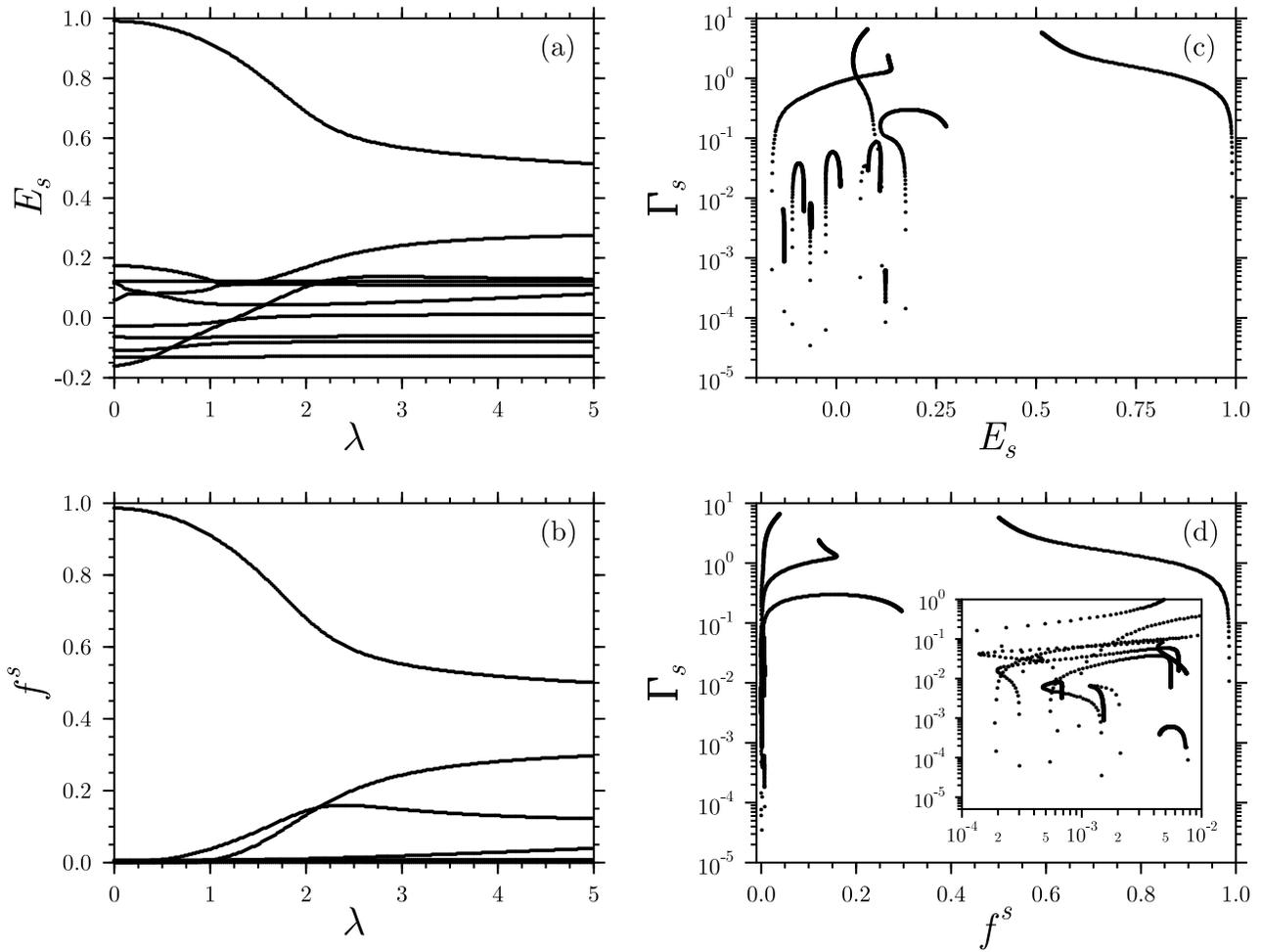}}}
\end{picture}
\unitlength 1bp
\caption{
The $\lambda$-dependence of resonance energies (a) and dipole strengths (b).
The logarithm of the widths versus energies (c) and dipole strengths (d) with
$\lambda$ varying from 0 to 5 in steps of 0.02. (Inset of (d) shows the
magnified region of small ${\tilde f}^{s}$ in the double log-scale). For
parameters see text.
}
%\label{fig1}
\end{figure}

\begin{figure}[t]
\unitlength 1cm
\begin{picture}(10,3)
\epsfxsize 15cm
\put(0,-5){\epsfbox{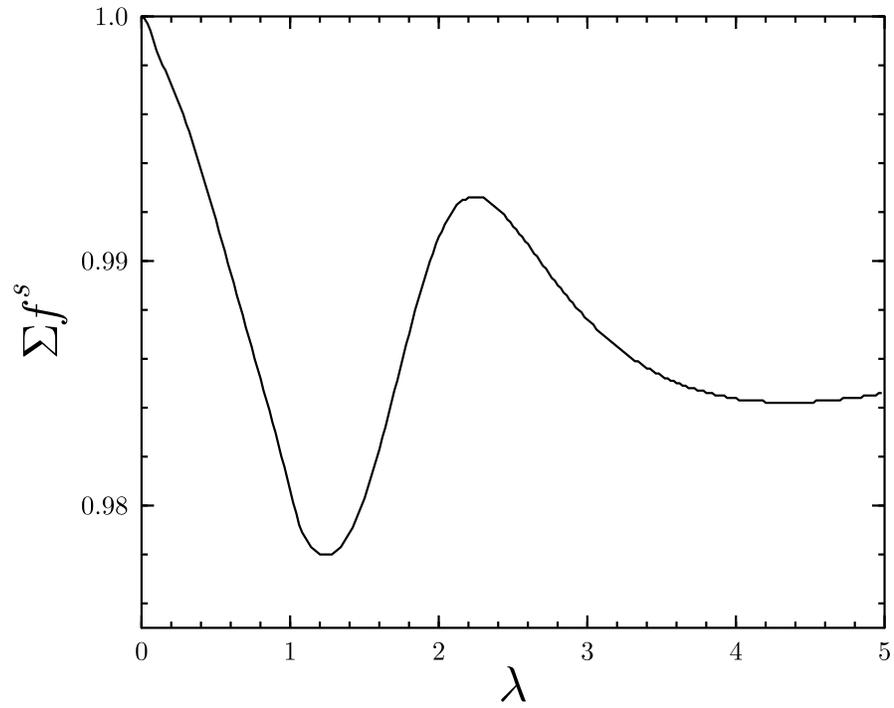}}
\end{picture}
\unitlength 1bp
\caption{
The sum of dipole strengths eqs. (\ref{resstr}, \ref{f-sp}) as a function
of $\lambda$.  
}
%\label{fig1}
\end{figure}

\end{document}